# Structural insight into $Co_{3-x}Mn_xTeO_6$; ($0 < x \leq 2$) solid solutions using Synchrotron X-ray diffraction and XANES

Harishchandra Singh[1], A. K. Sinha[1], Haranath Ghosh[1], M. N. Singh[1], Parasmani Rajput[2]

[1]Indus Synchrotron Utilization Division, Raja Ramanna Centre for Advanced Technology, Indore - 452013, India
[2]Atomic & Molecular Physics Division, Bhabha Atomic Research Centre, Trombay, Mumbai - 400085, India

**Abstract**

Structural investigation on $Co_{3-x}Mn_xTeO_6$; ($0 < x \leq 2$) solid solutions as a function of Mn concentration using Synchrotron X-ray diffraction (SXRD) and XANES measurements are presented. Phase diagram obtained from Rietveld Refinement on SXRD data as a function of Mn concentration indicates doping disproportionate mixing of both monoclinic (C2/c) and rhombohedral (R$\bar{3}$) structure (for $x < 0.5$) while only R$\bar{3}$ symmetry for $x \geq 0.5$. Furthermore, Rietveld analysis on SXRD data shows increase in lattice parameters and average transition metal − oxygen (Co/Mn-O) bond lengths for $x \geq 0.5$. Co and Mn K-edge XANES spectra reveal that both Co and Mn are in mixed oxidation state of $Co^{2+}/Mn^{2+}$ and $Co^{3+}/Mn^{3+}$. Relative ratio of $Co^{3+}/Co^{2+}$ and $Mn^{3+}/Mn^{2+}$ using Linear combination fit (LCF) decrease with increasing $x$ for $x \geq 0.5$. These structural and spectroscopic evidences are used to provide possible interpretation of the reported observations of maximum Neel Temperature ($T_N$) at $x \sim 0.5$ and enhancements of antiferro/ferro-magnetic interactions in $Co_{3-x}Mn_xTeO_6$ solid solutions.

# 1. Introduction

Apart from the understanding of origin/enhancement of the coupled interactions of ferroic orders, designing and finding new multiferroic materials are some of the frontier research activities in condensed matter physics [1]. These materials exhibit either coupling between electronic and magnetic orders (type II) or separate single order parameters (type I) [2]. In addition, controlling and coupling of various 'ferro or antiferro' magnetic orders at room temperature induced either in parent or doped material are of immense interest [3]. $Co_3TeO_6$ (CTO) is one of the recently found low symmetry (C$2/c$) type II multiferroic material at low temperatures, which shows complex magnetic structure with sequence of antiferromagnetic (AFM) transitions [4]. $Mn_3TeO_6$ (MTO), on the other hand, crystallizes in higher symmetry (R$\bar{3}$) type I multiferroic material, exhibits similar antiferromagnetic transitions at low temperature [5]. MTO and CTO show AFM transitions at 23K and 26K, respectively [4-5]. In contrast, Mn doping in CTO (~7.5 at. %) enhances the AFM transition temperature to ~ 40K; even when the transition temperatures of the end members are lower [6-7]. Mn substitutions on Co sites may have several effects (i) different size (atomic or ionic) of Mn as compared to Co can modify the effective orbital overlap between the transition metals and thus modifying the exchange integral, (ii) probable mixing of variable oxidation states of Mn/Co may also influence the effective magnetic moment and interactions and iii) chemical pressure may change electronic configuration, inter-atomic distances and local environment etc. As a result, changes in transition metal – oxygen (Co/Mn-O) bond lengths and their respective angles (TM-O-TM) along with variable oxidation states of Co and Mn greatly influence the magnetic properties of the compound. Important features of intermediate compound as compared to end members can be very exciting, like insulator-metal, antiferro-ferromagnetic transition and vice versa observed in LaMn$_{1-x}$Co$_x$O$_3$ etc. has lead us to investigate *Co$_3$-* $_x Mn_x TeO_6$ solid solutions [8-9]. To address issues related to these, we present room temperature Synchrotron X-ray diffraction (SXRD) and X-ray absorption near edge structure (XANES) studies for detailed structural information and its probable correlation to reported magnetism of $Co_{3-x}Mn_xTeO_6$ solid solutions.

Rietveld refinements based phase diagram obtained from SXRD data as a function of Mn concentrations indicate doping *disproportionate* mixing of both symmetry ($R\bar{3}$ and C*2/c*) for *x < 0.5* while only $R\bar{3}$ symmetry for *x ≥ 0.5*. Furthermore, average transition metal (TM) – oxygen (O) bond lengths vary varies with increase in Mn concentration and attains minimum value at x= 0.5. K edge XANES spectra at Co and Mn edges show mixed oxidation states of $Co^{2+}/Co^{3+}$ and $Mn^{2+}/Mn^{3+}$. Te $L_3$ edge XANES spectra show +VI oxidation state of Te in $Co_{3-x}Mn_xTeO_6$ solid solutions. Increase in TM - O bond lengths and decrease in charge ratio of $Co^{3+}/Co^{2+}$ and $Mn^{3+}/Mn^{2+}$ can influence the magnetic interaction between the transition metals and has been used to explain the reported magnetic behaviors [6-7]. Rest of the paper is organized as follows. Method of synthesis, equipment details and structural characterization of $Co_{3-x}Mn_xTeO_6$ solid solutions are described in the experimental section 2. We analyze the structural and XANES data in the results and discussion section 3. Detailed results of our structural and spectroscopic studies are presented in discussion section followed by conclusions section 4.

## 2. Experimental Method

Solid solutions of Mn doped cobalt tellurate have been synthesized via conventional solid state reaction route. The reactants used were cobalt oxide $Co_3O_4$ (Alpha Easer 99.7 %), $Mn_3O_4$ (obtained from high purity $MnO_2$: Alpha Easer 99.999%) and tellurium dioxide $TeO_2$ (Alpha Easer 99.99 %). $Mn_3O_4$ has been synthesized from $MnO_2$ by calcining $MnO_2$ at

~1100°C for 12 hrs [10]. We have prepared $Co_{3-x}Mn_xTeO_6$; ($x$ = 0.05, 0.1, 0.2, 0.3, 0.4, 0.5, 0.8, 1, 2). The ground oxide mixtures were heated first at 700°C for 10 hrs and then calcined at 800°C for more than 25 hrs in the second step. While performing the second step of growth, we increased the duration of calcinations from 25 hrs to 30 hrs with increase in Mn concentration ($x$ < 1). For $x$ = 1 and 2, the calcinations duration was increased to 48 hrs. For each step of calcination and sintering, the pellets were made by applying 2 tons of pressure. Compression in all the pellets for $x$ < 0.2 and no volume change for $x$ > 0.5 were observed, on calcinations. In the composition range $0.2 \leq x \leq 0.5$, volume expansion like behavior was observed. Anomalous behavior of volume changes could be due to the large variations in the densities of CTO (~ 1400 gm/cm$^3$) and MTO (720 gm/cm$^3$). With increase in Mn concentration, dark blue color of pure CTO changed to shining blue for $0.2 \leq x \leq 0.5$ and light blue for $0.2 > x > 0.5$. Characterizations of synthesized samples were done using SXRD and XANES. Both the above measurements were performed at angle dispersive X-ray diffraction (ADXRD) beamline (BL-12) on Indus-2 synchrotron source, RRCAT India [11]. Beamline consists of Si (111) based double crystal monochromator and two experimental stations namely a six circle diffractometer with a scintillation point detector and Mar 345 Image Plate area detector. Present SXRD measurements were carried out using the area detector. The X-ray wavelength used for the present study was accurately calibrated by doing X-ray diffraction on the lanthanum hexaborate (LaB$_6$), NIST standard on Image Plate. Wavelength used for the present measurements were 0.9528 Å and 0.7219 Å. The two dimensional patterns were integrated using fit2D programme [12]. The refinement of the structure parameters from the diffraction patterns was done using the Rietveld method employing the Fullprof program [13]. XANES measurements were carried out in transmission mode (for Co K –edge) and fluorescence mode (for Mn K –edge) at room temperature. The photon energy were calibrated by the Co/Mn K-edge XANES spectra of standard Co/Mn metal at 7.709 / 6.539 keV. The fluorescence XANES spectra were recorded using vortex energy dispersive detector ((VORTEX-EX). Te $L_3$ edge XANES spectra were recorded at Scanning EXAFS Beamline (BL-9) at the Indus-2.

## 3. Results and discussion

### 3.1. Synchrotron X-ray diffraction studies

For structural verification, SXRD measurements have been performed on all the compositions of $Co_{3-x}Mn_xTeO_6$ ($x$ = 0.05, 0.1, 0.2, 0.3, 0.4, 0.5, 0.8, 1.0, 2.0). All the data have been collected at room temperature. Rietveld refinement on the SXRD data has been used for the determination of space group symmetry and lattice parameter variations as a function of Mn concentration. Fig. 1 shows SXRD patterns for the all the samples with $x$ = 0.05, 0.1, 0.2, 0.3, 0.4, 0.5, 0.8, 1.0, 2.0. The figure 1 clearly indicates phase transition at $x$ = 0.5. A closer look at the SXRD pattern reveals that the composition with $x$ = 0.05, 0.1, 0.2 (even with $x$ = 0.3, 0.4) has some extra peaks as compared to $x$ = 0.5, 0.8, 1.0 and 2.0. On comparing these SXRD patterns to that of the corresponding end members i.e., CTO ($x$ = 0) and MTO ($x$ = 3), one can have the clear idea regarding space group symmetry possessed by these compositions. Ceramic CTO structure possesses monoclinic (space group C$2/c$) symmetry whereas, MTO exhibits rhombohedral (R$\bar{3}$) symmetry [5, 14-15]. While doping different concentrations of Mn in CTO, we have found that for lower doping concentration i.e. for $x$ < 0.5, mixed phases of CTO (C$2/c$) and MTO (R$\bar{3}$) co-exist. Crystal structure of both the CTO and MTO are shown in Fig. 2 [14-15, 5]. Two phase Rietveld refinement for samples with $x$ = 0.05, 0.1, 0.2, 0.3 and 0.4 has been used. Samples with $x$ = 0.05 and 0.1 show very weak reflections of MTO (R$\bar{3}$) phase, which indicates that majority of the material is in the C$2/c$ phase. The samples with $x$ = 0.2 and 0.3 possess R$\bar{3}$ phase dominantly. This is quite unusual because the samples still have majority of Co (2.8) as compared to Mn (0.2) for $x = 0.2$. Two phases Rietveld refinement analysis shows the presence of C2/c phase of ~ 5%, ~3% for $x = 0.3$ and 0.4 respectively [16]. In the literature, it has been reported that for $x = 0.3$ observed magnetization has small kink at ~ 26K (for C2/c phase) apart from the major transition at 40K, which is not explained adequately, so far [7]. Changes in lattice parameters of C2/c (CTO) and R$\bar{3}$ (MTO) phases are shown in Figure 3 for $x < 0.5$. From the Figs. 3 (a) and 3 (b), it has been observed that lattice parameters $a$ and $b$ increases whereas $c$ and $\beta$ decreases, with increasing $x$. Typical error bar in all the bond lengths and lattice parameters is estimated to be ~ $2\times10^{-4}$ Å. Therefore, we do not show the error bar in the respective figures but the error bars are within the dimensions of the symbols. On the other hand, lattice parameters ($a = b$ and c) of R$\bar{3}$ (MTO) phases show parabolic like behavior, Fig. 3 (c). This may be due to the lower atomic radii of Mn as compared to Co [17]. However, shrinkage/expansion of lattice parameters for both the phases with increase in Mn concentration (for $x < 0.5$) is anisotropic in nature (Fig. 3).

Furthermore, Rietveld refinements on $x \geq 0.5$ reveal solid solution with symmetry same as that of MTO. The experimental, calculated, and its difference powder x-ray diffraction profiles are shown in Fig. 4 for $x = 1.0$. All the Rietveld refinement parameters along with quality factors are included in table 1 for the compositions $x = 0.5$, 1.0 and 2.0. Rietveld refinements on all the samples for $x \geq 0.5$ indicate almost linear increase in lattice parameters and volume of R$\bar{3}$ unit cell with increase in $x$ which approaches towards lattice parameter of MTO (shown in Fig. 5). This indicates the relaxation of R$\bar{3}$ unit cell with increase in Mn concentration. The reported values of lattice parameters, in case of CTO (C2/c), are a = 14.8061(2) Å, b= 8.8406(3) Å, c= 10.3455(1) Å, β=94.819(4)° whereas, the same for MTO (R$\bar{3}$) are a = b = 8.8675(3) Å and c= 10.6731(2) Å [14-15, 5]. During the Rietveld refinement, octahedral

Co/Mn cations sit at 18*f* *Wyckoff* site, assuming random distribution of Co/Mn. Both the Te$_1$ and Te$_2$ occupy octahedra configuration 3*b* and 3*a* respectively. The Mn/CoO$_6$ octahedron is considerably distorted which is reflected in the variation of the transition metal-oxygen (Mn/Co-O) bond distances. On the other hand, Te-O octahedra (TeO6) are less distorted. The TM-O bond lengths (Mn/Co-O) vary from 2.0154(2) Å to 2.2832(2) Å for *x* = 0.5 and 2.0551(2) Å to 2.3463(2) Å for *x* = 2. Mn/CoO6 octahedra (inferred through bond lengths) are contracted as compared to the reported one [7]. Representative octahedral distortion (for *x* = 0.5 and *x* = 2.0) through a particular transition metal (shown by arrow in Fig. 2 (b)) in oxygen octahedral environments has been shown in Fig. 6. These results will further compliment the observations of Co$^{3+}$ and Mn$^{3+}$ using XANES measurements (to be discussed in next section) as Co$^{3+}$/Mn$^{3+}$ prefer smaller site as compared to Co$^{2+}$/Mn$^{2+}$ [18]. The corresponding average bond length of Mn/Co-O along with lattice parameters and volume for $R\bar{3}$ is shown in Fig. 5 (for *x* ≥ 0.5), where values corresponding to *x* = 3.0 has been incorporated from Ivanov et al. for comparisons [7]. The transition metal (TM) - transition metal bond lengths (Mn/Co-Mn/Co) range from 3.2171(2) Å to 4.4522(3) Å. This may indicate that the exchange interaction is favored through nonmagnetic oxygen anions. Smaller TM-O-TM lengths increase the strength of super-exchange interaction which in turn increases the Neel temperature (T$_N$). Structural visualization and bond lengths/angles analysis has been done using Material Studio 6.1 CASTEP package and VESTA software [15]. Fig. 7 shows the phase diagram obtained by Rietveld refinements (two phases for *x* < 0.5 and single phase for *x* ≥ 0.5) which indicate doping disproportionate concentration of $R\bar{3}$ and C2/*c* phases below *x* < 0.5. Furthermore, we did not observe any drastic change in any positional coordinates with increasing Mn concentration from room temperature SXRD data (see table 1 for *x* = 0.5, 1.0 and 2.0). The average oxidation state and the corresponding spin state (high spin state for Mn and low/intermediate/high spin state for Co) will also influence the magnetic behavior of the system. To study the average oxidation state of TM, we have performed Mn and Co K-edge XANES measurements as a function of Mn concentration. XANES results are discussed in detail in the next section.

### 3.2. X-ray absorption near edge structure (XANES) studies

XANES is an element specific spectroscopic tool which is extremely sensitive to the oxidation state, local coordination and hybridization effect of orbital of the elements present in the sample [19]. The edge step normalized XANES spectra at Co/Mn K-edge of $Co_{3-x}Mn_xTeO_6$ series as well as standard references, CoO and $CoF_3$ for Co and $MnCl_2$, $Mn_3O_4$, $Mn_2O_3$ for Mn of known oxidation states are presented here. In both the cases, photon energy has been calibrated by measuring absorption through XANES of respective metal foil (Co/Mn metal). Normalized XANES spectra for $Co_{3-x}Mn_xTeO_6$ series at Co K edge is shown in Figure. 8. XANES shapes are similar for all the composition studied, exhibiting a structured pre-edge region and the dominant peak called white line peak together with rising main edge [16]. Herein, only main edge part of the XANES spectra has been emphasized. The analysis of the Co K- edge shifts have been performed using main edge energies of the standard reference samples like CoO and $CoF_3$ for formal valence states $Co^{2+}$ and $Co^{3+}$, respectively. The main edge energy corresponding to Co K edge for the solid solutions of $Co_{3-x}Mn_xTeO_6$ is intermediate between those of CoO and $CoF_3$ (main panel, Fig. 8). As the energy position of the maxima of first derivative decreases with increasing $x$ (right inset of Fig. 8), samples oxidation states decreases gradually from $Co^{3+}$ to $Co^{2+}$. This suggests that the average valence state shifts towards $Co^{2+}$ with increasing Mn content. We calculate relative concentration of mixed valence in a sample by noting their energy positions and using a simple linear combination formula, "Main edge energy positions of sample's XANES spectra = {Energy position of $1^{st}$ standard × $y$ + Energy position of $2^{nd}$ standard × (1-$y$)} /100", where *y* is the calculated concentration of 1$^{st}$ standard [15]. Energy position of main edge in XANES spectra of a sample may be determined either as the energy corresponding to ~ 0.5 absorption or maximum energy value of a first order differentiated spectrum. Above formula has been derived by assuming a linear dependence of the chemical shift on the average valence by which one can obtain quantitative information on the valence state in a sample. Based on the above idea, quantitative phase composition analysis by Linear Combination Fitting (LCF) method on XANES data, using software Athena, has also been done [16, 20]. This approach used in XANES experiments is based on statistical goodness-of-fit criteria [20]. We have reported recently that the compositions obtained from LCF on mixed phase oxide samples matches well with that obtained from Rietveld refinements of SXRD data [16]. As we have already confirmed (above section) that the present solid solution shows mix phase (C2/*c* and R$\bar{3}$) behavior for x < 0.5 and single phase (R$\bar{3}$) for x ≥ 0.5. In this section, we will not be emphasizing x < 0.5 part but mainly x ≥ 0.5. Quantitative analysis using LCF and linear combination formula give the information regarding relative concentration of $Co^{2+}$ and $Co^{3+}$ in the corresponding solid solution. For example, as shown in left inset of Fig. 8, LCF fitting on sample with *x* = 0.2 gives ~ 37% $Co^{3+}$ and ~ 63% $Co^{2+}$ with ~ 2.37 as an average valence state, in mix phase structure (*x* < 0.5). Further, in single phase (*x* ≥ 0.5), relative ratio of $Co^{3+}$/$Co^{2+}$ decreases with *x* from ~ 0.4 for *x* = 0.5 to ~ 0.2 for *x* = 2.0 and show maximum charge disproportion at *x* = 0.5 (see left inset of Fig. 8). Similar to Co K edge, edge step normalized XANES spectra for Mn K edge (6539 eV) is shown in Figure. 9. We have not shown XANES spectra for all the samples for brevity. Shift in the main edge (rising edge) with change in doping concentration for all the samples is shown in the left inset of the Fig. 9. The right inset shows LCF fit for *x* = 0.1. It is interesting to note that the main edge energy for all the solid solution of *Co$_{3-x}$Mn$_x$TeO$_6$* are shifted towards $MnCl_2$ ($Mn^{2+}$) from $Mn_2O_3$ ($Mn^{3+}$) with increasing *x*, however the shift shows $Mn^{2+}$ valence dominantly (right inset of Fig. 9). This suggests that with increasing Mn concentration, oxidation state approaches 2+ states. Qualitative analysis of the Mn K-edge shifts has been performed using main edge energies of standards reference samples such as $MnCl_2$, $Mn_3O_4$ and $Mn_2O_3$ for formal valence states $Mn^{2+}$, $Mn^{2.67+}$ and $Mn^{3+}$, respectively. Quantitative analysis, same as above (Co K edge), for example sample with $x = 0.1$ (left inset of Fig. 9) shows ~ 70% $Mn^{2+}$, ~ 30% $Mn^{3+}$ and thus ~ 2.30 as an average valence state of Mn. Moreover, with increase in $x$ value, the ratio of $Mn^{3+}$/$Mn^{2+}$ decreases from ~ 0.26 for $x = 0.5$ to ~ 0.13 for $x = 2.0$ and show maximum at $x$ ~ 0.5 in the concentration range $0.5 < x < 2$. Though, $Co^{3+}$ concentration is higher as compared to $Mn^{3+}$ for the corresponding solid solution. Based on XANES measurements of *$Co_{3-x}$ $Mn_xTeO_6$* ($x = 0.5, 1.0$ and $2.0$) at the Te $L_3$ edge, we show that the Te oxidation state in these samples is +VI. Te $L_3$ edge XANES spectra for these samples coincide (main edge part) with that of the standard samples containing $Te^{+IV}$ state, shown in Fig.10. Comparison of these particular samples with two standard samples ($Te^{+IV}$ and $Te^{+VI}$) confirm the +VI oxidation state of Te in $Co_{3-x} Mn_xTeO_6$. This result of $Te^{+VI}$ oxidation state is very much consistent with other reports [21]. XANES observations on Mn, Co and Te edge indicate either oxygen excess or cations vacancy in the samples [22-23].

Mathew et al. and Ivanov et al. have recently reported that the change in magnetic transition temperature with Mn concentration doesn't have structural origin [6-7]. On the contrary, present SXRD and XANES results suggest possible explanation for the variations in the magnetic transition temperature with change in Mn concentration. Mn doping at Co site and/or vice versa may affect the structure either via transition metal ion's size or by chemical pressure introduced into the lattice. In *$Co_{3-x}Mn_xTeO_6$*, these changes are reflected in the Mn/Co-O bond distances, though not in Mn/Co-O-Mn/Co bond angles so far. In addition, the average Co/Mn-O bond distances which show a linear increase (for $x \geq 0.5$, Fig. 5) with increase in $x$, we report for the first time, that transition metals (Co/Mn) show mix valency. Mn/Co mixed valency observed through XANES would induce different exchange interaction in the sense that super exchange through like spins ($TM^{2+/3+}$ - O - $TM^{2+/3+}$) and double exchange between unlike spins ($TM^{2+/3+}$ - O - $TM^{3+/2+}$) [24-27]. Generally, $Co^{2+}/Mn^{3+}$ state is more electrochemically favored than the $Co^{3+}/Mn^{2+}$ state, which supports the reduction of $Co^{3+}$ to $Co^{2+}$ rather than reduction of $Mn^{3+}$ to $Mn^{2+}$ [28]. However, in our case Mn/Co cations exist dominantly in $Mn^{2+}/Co^{2+}$ oxidation state. Due to completely antiferromagnetic nature of magnetic order in CTO and MTO, it is likely that the same magnetic order exists in these solid solutions, as well [4-5]. Though weak ferromagnetism in the single crystal of CTO has been suggested via spin canting where only one magnetic $Co^{2+}$ ion is proposed [4]. The same mechanism of ferromagnetism may also be possible through the double exchange interaction ($TM^{2+/3+}$ – O – $TM^{3+/2+}$) in these solid solution, because our XANES data clearly show the presence of $Co^{3+}$ in addition to $Co^{2+}$ [24-27].

We conclude our discussion and provide a probable explanation to the enhancement in effective magnetic moment and $T_N$ (for $x \sim 0.5$) of the corresponding solid solution, based on our SXRD and XANES measurements [6]. Fig. 5 (c) indicate that the average TM - O bond length is minimum for $x \sim 0.5$, where the corresponding charge ratios ($Co^{3+}/Co^{2+}$ and $Mn^{3+}/Mn^{2+}$) are maximum compared to higher $x$ values. Approximately, the same value of $x$ corresponds to the maxima in $T_N$ [6-7]. The ionic radii of $Co^{2+} \sim 0.75$Å (HS) and $Co^{3+} \sim 0.55$Å (LS)/0.61Å (HS) where as that for $Mn^{2+} \sim 0.83$Å (HS) and $Mn^{3+} \sim 0.58$A (LS)/0.64Å (HS), indicates $Mn^{3+}$ would favor $Co^{3+}$ and $Mn^{2+}$ to $Co^{2+}$ (where HS stands for high spin and LS for low spin) [17]. This may further explain the obtained lattice parameters and average bond distance variations for $x \geq 0.5$. Recently, we have studied the *coexistence of $Co^{2+}/Co^{3+}$ in ceramic $Co_3TeO_6$* and found concentrations of $Co^{2+}$, $Co^{3+} \sim 60$ and $40\%$, respectively [15].

We have considered low/high spin state for $Co^{3+}$ and high spin state for $Co^{2+}$ that provides favorable ground state based on our *first principle calculations* on CTO [15]. Similarly, one may consider the possibility of high spin state of the corresponding Mn in the present solid solution (Mn favors generally high spin). As in the high spin state, $Mn^{2+}$ with S=5/2 leads to magnetic moment of 5.92 $\mu_B$ per Mn ion whereas $Co^{2+}$ with S=3/2 to 4.89 $\mu_B$ per Co ion. Furthermore, as per our XANES studies, with increase in *x* value the main edge shift (discussed above in XANES section) infers $Mn^{2+}$ dominantly on Mn doping. This naturally leads to enhanced spin moment per site with increase in *x*. Along with above, our observation of minimum < TM-O-TM bond > signifies to the possibility of larger super-exchange interaction at *x* = 0.5. Therefore, enhancements in the super-exchange interaction due to minimum average TM - O (Mn/Co-O) bond lengths, where the relative ratio of $Co^{3+}/Co^{2+}$ and $Mn^{3+}/Mn^{2+}$ are also found to be maximum, results maximum in $T_N$ at *x* ~ 0.5. To obtain a complete insight in to the magnetic and structural correlation, magnetization and first principle studies will be discussed elsewhere.

## 4. Conclusions

We report structural investigation on *$Co_{3-x}Mn_xTeO_6$* (0 < *x* ≤ 2) solid solutions as a function of Mn concentration using SXRD and XANES studies. Detailed Rietveld refinements show $R\bar{3}$ structure for *x* ≥ 0.5 whereas mixture of both C2/c and $R\bar{3}$ structure for *x* < 0.5. For *x* ≥ 0.5, increase in lattice parameters and average TM - O bond distances with increasing in Mn concentration naturally corroborate with the reported magnetic behaviour. Analysis of XANES spectra at Co and Mn K-edges show mixed oxidation states of $Co^{2+}/Co^{3+}$ and $Mn^{2+}/Mn^{3+}$. Quantitative analysis of oxidation state using Linear combination fit (LCF) indicates gradual increase in $Co^{2+}$ and $Mn^{2+}$ with increase in Mn concentration. Relative ratios of $Co^{3+}/Co^{2+}$ and $Mn^{3+}/Mn^{2+}$ are found to be maximum at around *x* ~ 0.5, where average Co/Mn-O bond length is minimum. At approximately the same concentration, Mathew et al., found maximum $T_N$. Together with these observations, we provide a basis for the cause of enhancements of antiferro/ferro-magnetic interaction in $Co_{3-x}Mn_xTeO_6$ solid solutions using structural and spectroscopic evidences. Therefore, through these studies influence of structural aspects on magnetic, electronic properties and their possible coupling are established.

**Acknowledgement**

Authors acknowledge Dr. S. K. Deb and Dr. G. S. Lodha for their support and encouragement. Harishchandra Singh gratefully acknowledges Homi Bhabha National Institute, India for providing Research fellowship.

**Table and Figure Captions:**

**Table. 1**. Results of the Rietveld refinements of the crystal structure of the Mn doped CTO samples (here $x$ = 0.5, 1.0 and 2.0 only) at room temperature using Synchrotron X-ray powder diffraction data using $R\bar{3}$ space group.

**Figure. 1.** Synchrotron X-ray diffraction patterns of $x$ = 0.05, 0.1, 0.2, 0.3, 0.4, 0.5, 0.8, 1 and 2 compositions possessed either mixed ($C2/c$ and $R\bar{3}$) symmetry (0.05, 0.1, 0.2, 0.3, 0.4) or pure $R\bar{3}$ symmetry (0.5, 0.8, 1.0, 2.0).

**Figure.2. a**) CTO, **b**) MTO crystal structure with $C2/c$ and $R\bar{3}$ symmetry, where red dot color represents the oxygen atoms, blue and magenta dots for Co and Mn atoms while the large dark yellow colors show the Te atoms along with the corresponding orientations. Arrow in Fig. 1 (b) shows a particular transition metal in oxygen environment.

**Figure. 3. a, b**) Variations of lattice parameter of $C2/c$ phase and (**c**) $R\bar{3}$ phase as a function of Mn concentrations for $x$ < 0.5, error bars are within the symbols (see text).

**Figure. 4.** Rietveld refinement of SXRD patterns for $x$ = 1.0 indicates the pure $R\bar{3}$ phase. Red circle represent raw data, black solid line the Rietveld fit, blue vertical bar the Bragg reflections and zigzag magenta line the difference between observed and calculated intensity.

**Figure. 5. a**) Variations of $R\bar{3}$ phase volume, **b**) lattice parameters and **c**) average transition metal-oxygen bond distances as a function of Mn concentration for $x \geq 0.5$ along with $x$ = 0.3 within the error bars (see text).

**Figure. 6.** Octahedra distortion for $x$ = 0.5 and 2.0, obtained from Rietveld refined structure, clearly showing bond length variations for $x$ = 0.5 to $x$ = 2.0. This particular transition metal has been shown by arrow in Fig. 2 (b).

**Figure. 7.** Phase diagram obtained from Rietveld Refinements on the SXRD data as a function of Mn concentration indicates mixed phases of C$2/c$ and R$\bar{3}$ < 0.5 while pure R$\bar{3}$ phase for ≥ 0.5.

**Figure. 8.** Edge step normalized XANES spectra at Co K edge, which show gradual shift of main edge energy indicating increase of Co$^{2+}$ concentration (right inset) with increase in Mn concentration using maxima in the first derivative of normalized absorption and the corresponding charge proportion (left inset).

**Figure. 9.** Edge step normalized XANES spectra at Mn K edge for $x$ = 0.1 and the corresponding LCF fit (right inset) along with gradual shift of main edge (raising edge) energy, via maxima in the first derivative which indicates increase of Mn$^{2+}$ concentration with increase in Mn concentration (left inset).

**Figure. 10**. Te L$_3$ edge XANES spectra of $x$ = 0.5 sample (main panel), showing +VI oxidation state of Te in the corresponding solid solution. Here we have used two standards reference samples, TeO$_2$ for Te$^{+IV}$ state and Te(OH)$_6$ for Te$^{+VI}$ oxidation states respectively. Inset show 1$^{st}$ derivative for $x$ = 0.5, 1.0 and 2.0 as a function of energy, indicating the same +VI oxidation state of Te in all the samples.

**Table 1**

| Phase | | $x = 0.5$ | $x = 1.0$ | $x = 2.0$ |
|---|---|---|---|---|
| a/Å | | 8.6398(3) | 8.7117(1) | 8.7964(3) |
| c/Å | | 10.4934(2) | 10.5628(2) | 10.6276(3) |
| Mn/Co | x/a | 0.0413(1) | 0.0421(3) | 0.0409(1) |
| | y/b | 0.2667(2) | 0.26557(4) | 0.2655(3) |
| | z/c | 0.2101(3) | 0.21215(3) | 0.2119(2) |
| | B/A2 | 0.64(4) | 0.18(3) | 0.06(2) |
| Te1 | x/a | 0.0 | 0.0 | 0.0 |
| | y/b | 0.0 | 0.0 | 0.0 |
| | z/c | 0.5 | 0.5 | 0.5 |
| | B | 0.33(2) | 0.06(3) | 0.03(4) |
| Te2 | x/a | 0.0 | 0.0 | 0.0 |
| | y/b | 0.0 | 0.0 | 0.0 |
| | z/c | 0.0 | 0.0 | 0.0 |
| | B | 0.77(2) | 0.03(2) | 0.12(4) |
| O1 | x/a | 0.0323(3) | 0.0295(1) | 0.0289(2) |
| | y/b | 0.2077(1) | 0.1971(4) | 0.1999(5) |
| | z/c | 0.3983(1) | 0.40034(2) | 0.3996(2) |
| | B | 0.37(2) | 0.02(3) | 0.38(3) |
| O2 | x/a | 0.1883(1) | 0.1859(1) | 0.1868(4) |
| | y/b | 0.1667(2) | 0.1614(1) | 0.1601(1) |
| | z/c | 0.1134(1) | 0.1149(3) | 0.1135(3) |
| | B | 0.78(3) | 0.21(2) | 0.23(4) |
| Rp | | 2.94 | 2.14 | 3.72 |
| Rwp | | 3.85 | 2.85 | 4.76 |
| Rb | | 4.01 | 4.14 | 3.21 |
| $\chi^2$ | | 0.921 | 0.474 | 2.19 |

**Fig. 1**

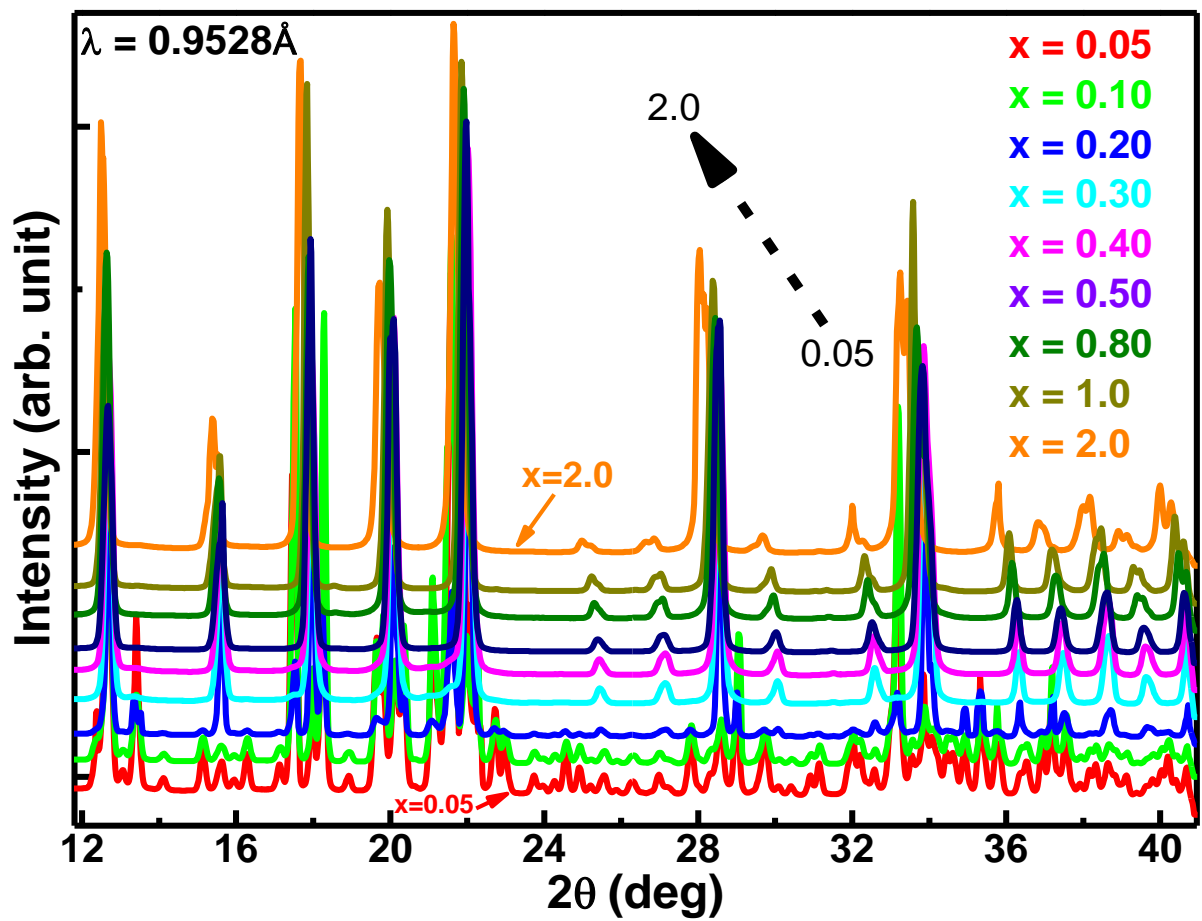

**Fig. 2**

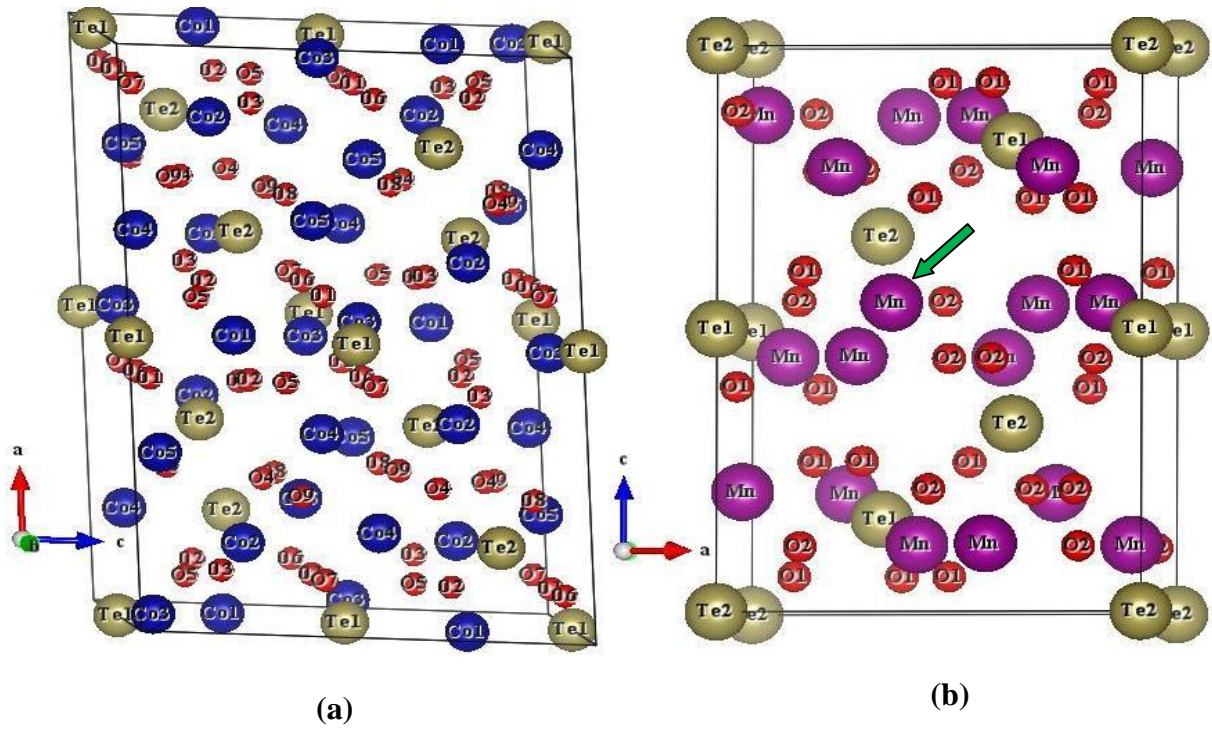

(a)           (b)

**Fig. 3**

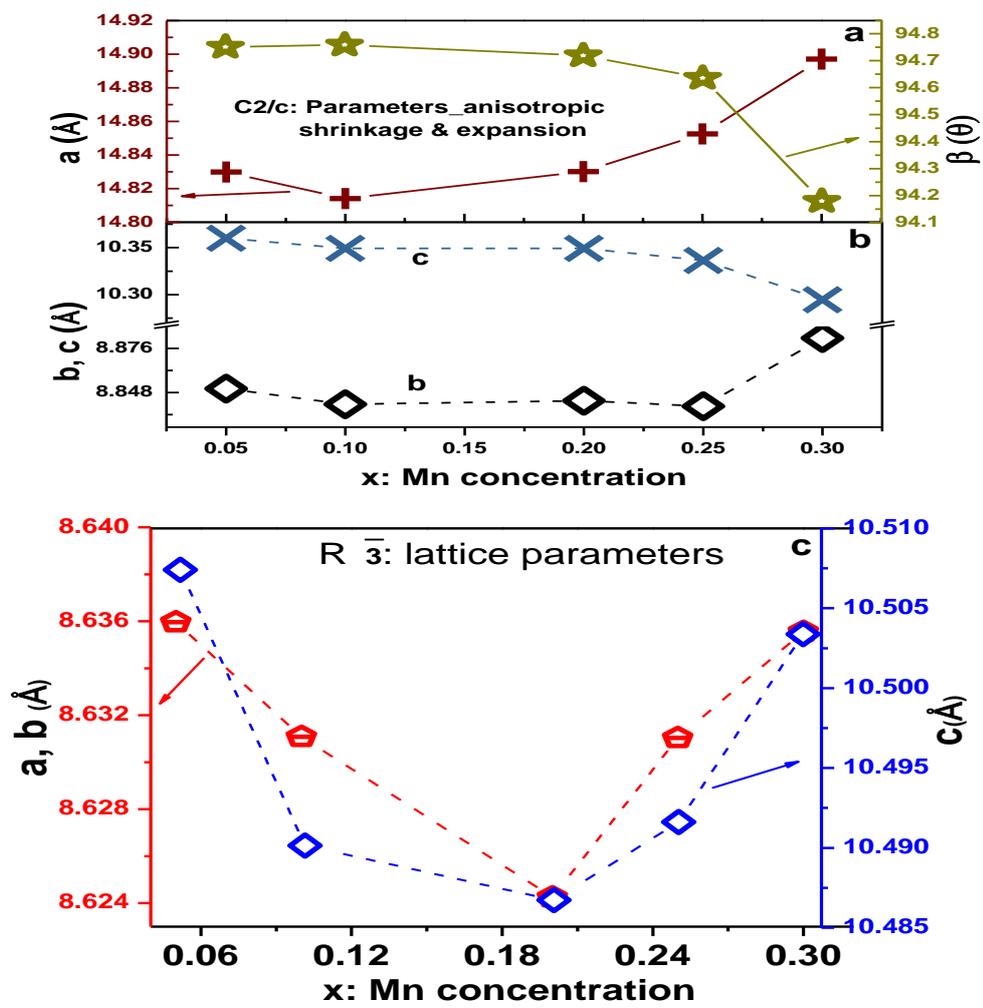

Fig.4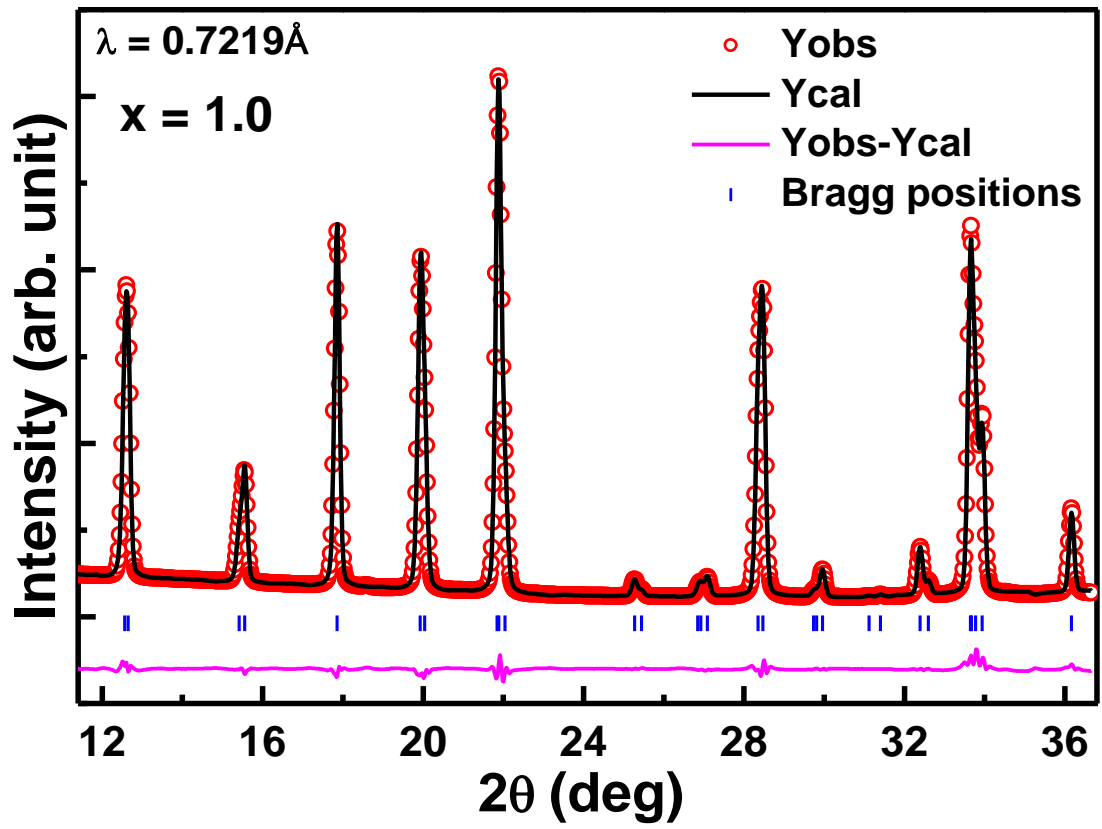

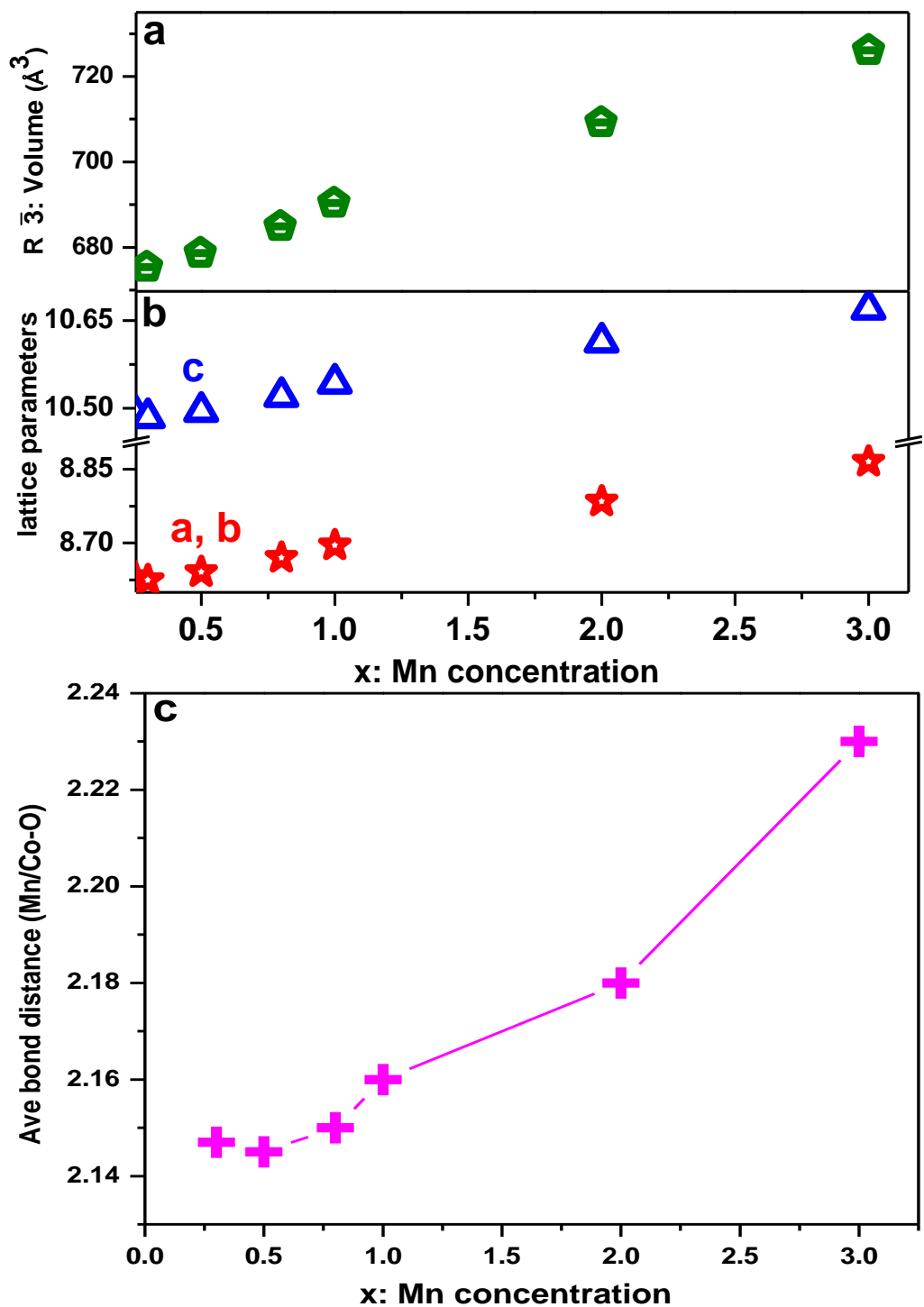

**Fig. 6**

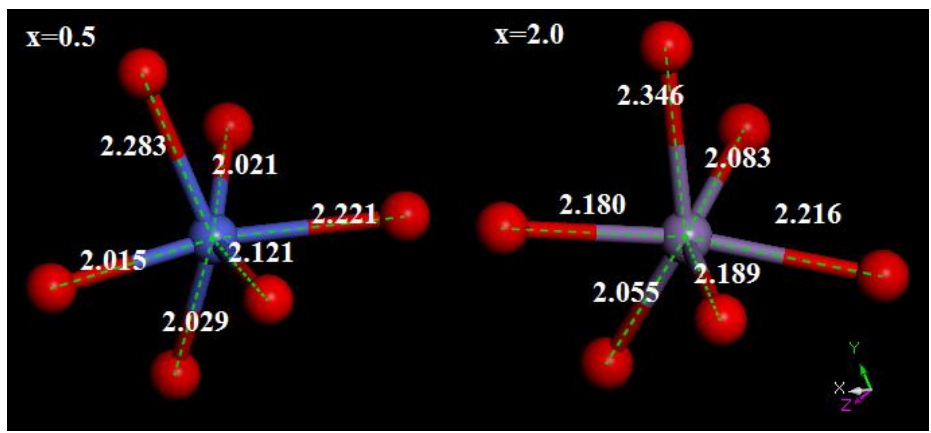

**Fig. 7**

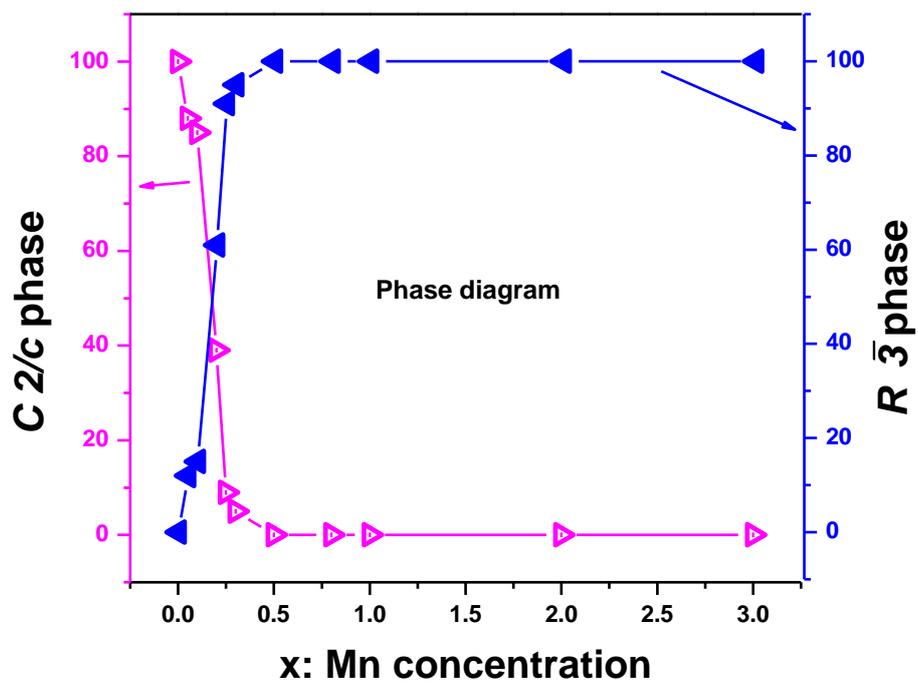

**Fig.8**

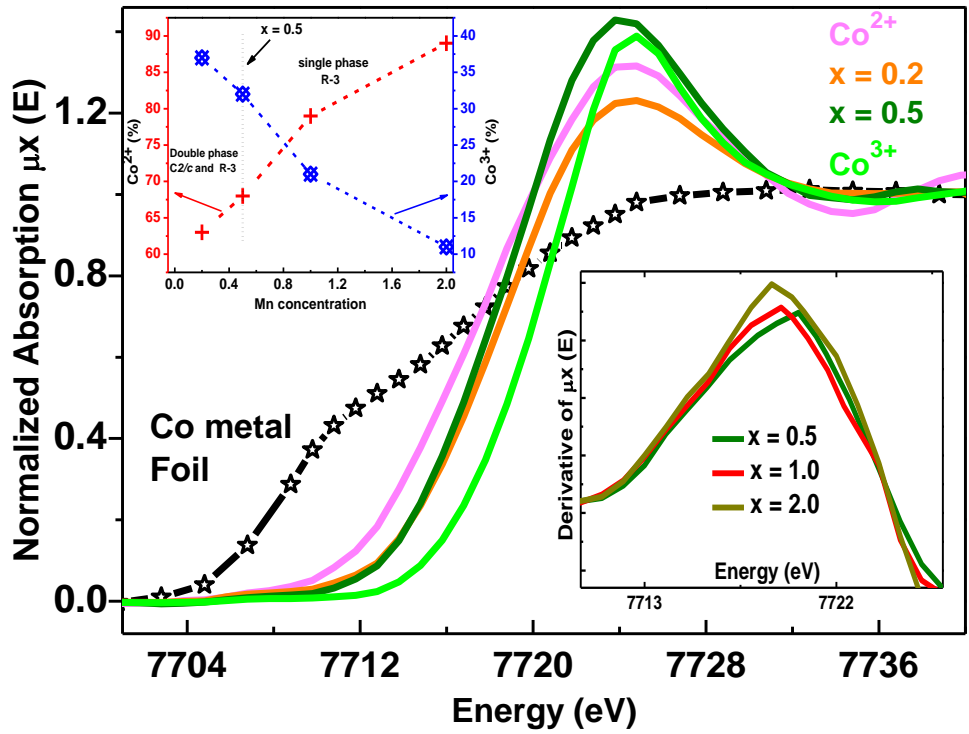

**Fig.9**

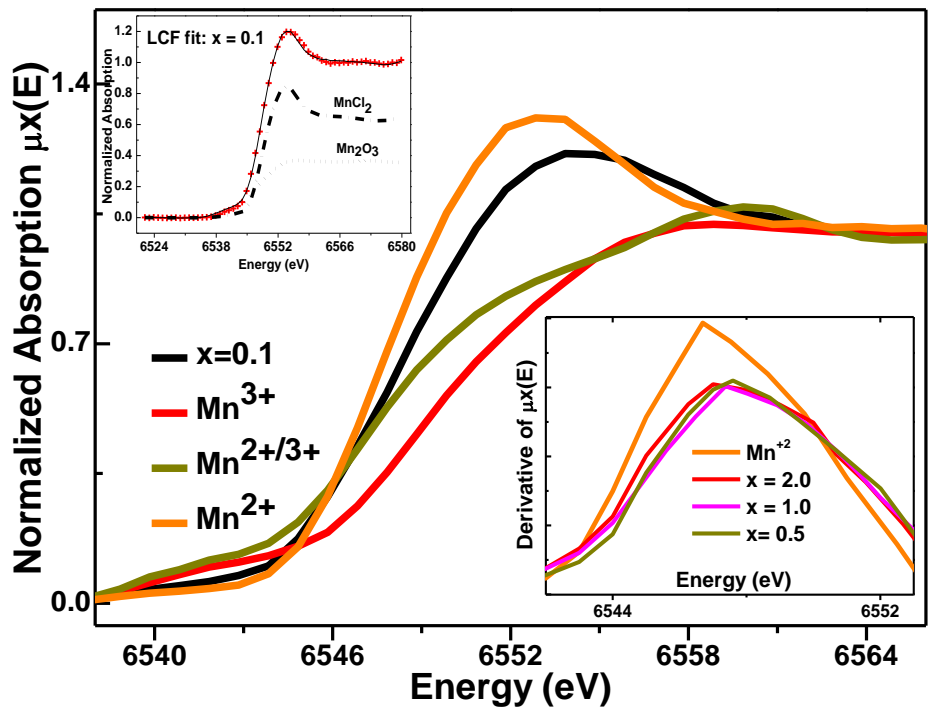

**Fig. 10**

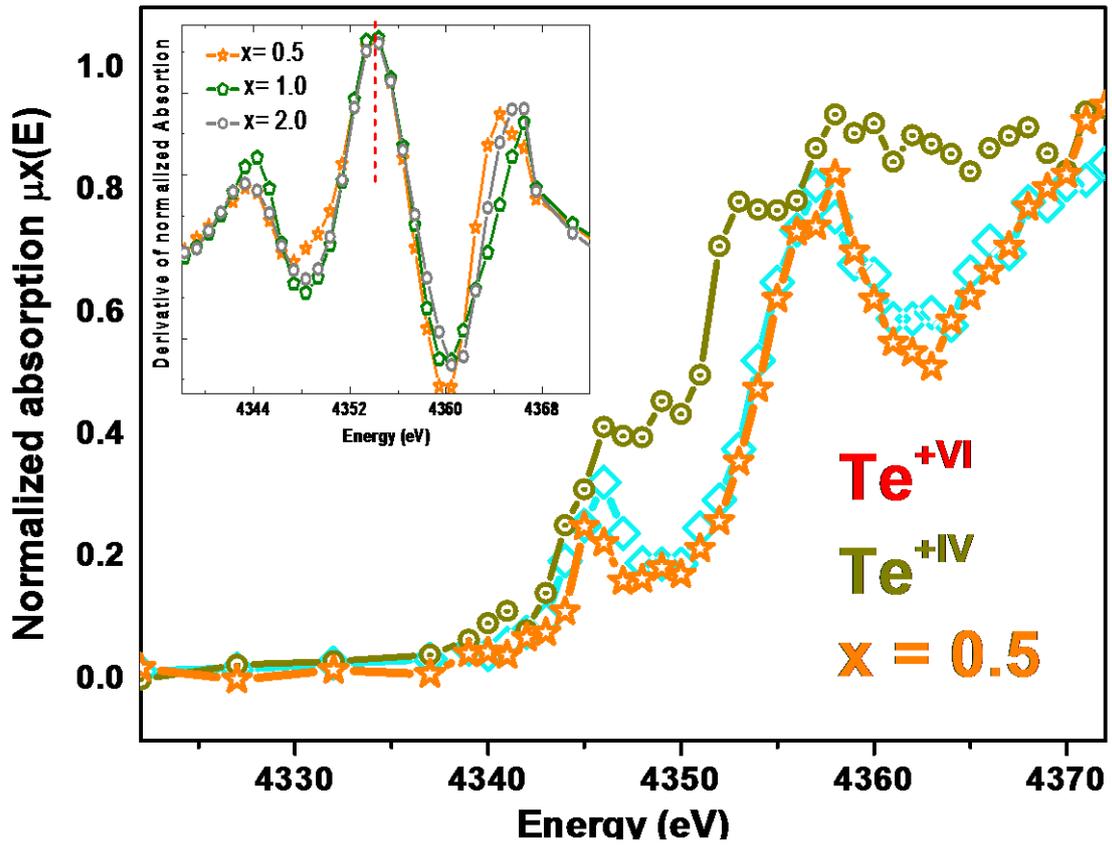